\documentclass[9pt]{elife}

\usepackage{lipsum} 
\usepackage[version=4]{mhchem}
\usepackage{siunitx}
\usepackage[table]{xcolor}
\usepackage{makecell}
\usepackage{placeins}

\definecolor{noColor}{HTML}{F4CCCC}
\definecolor{yesColor}{HTML}{D9EAD3}
\definecolor{cpcColor}{HTML}{D9D2E9}
\definecolor{mseColor}{HTML}{FFF2CC}
\DeclareSIUnit\Molar{M}

\title{Local learning is all you need: brain-aligned predictive coding networks}

\author[1,2*]{Dirk Gütlin}
\author[1,3\authfn{1}]{Ryszard Auksztulewicz}
\author[1,2\authfn{1}]{Adrien Doerig}

\affil[1]{Department of Education and Psychology, Freie Universität Berlin, Berlin, Germany}
\affil[2]{Bernstein Center for Computational Neuroscience Berlin, Berlin, Germany}
\affil[3]{Department of Neuropsychology and Psychopharmacology, Faculty of Psychology and Neuroscience, Maastricht University, Maastricht, Netherlands}

\contrib[\authfn{1}]{These authors share last authorship}

\corr{dirk.guetlin@fu-berlin.de}{DG} 

\begin{document}

\maketitle

\begin{abstract}
Predictive Coding (PC) is an influential account of how brain representations emerge through hierarchical, local, self-supervised prediction-based learning, without requiring globally propagated error signals. Despite extensive evidence supporting PC, a striking paradox persists: computational models built on these biologically grounded principles have consistently failed to match state-of-the-art Deep Neural Networks in performance and brain-alignment. Here, we show that this paradox can be overcome. We introduce Predictive Brain Networks (PBNs), a hierarchical recurrent framework that implements local PC-based learning and achieves representational alignment with the human ventral visual stream matching or exceeding established vision models. Importantly, PBN top-down predictions encode unique, brain-aligned information not captured during bottom-up feature extraction, with this contribution increasing in locally trained variants. Further, PBNs substantially outperform existing PC-inspired networks on downstream classification tasks, reproduce key signatures of PC (with higher-level predictions shaping lower-level visual responses as observed in the primate cortex), and support uncertainty estimation. PBNs therefore close a longstanding gap in computational neuroscience by combining two properties that have been largely separated: the brain-inspired learning mechanisms of predictive coding and the representational power of modern deep networks. They provide a scalable framework for studying biologically grounded learning without sacrificing the capacity to produce rich neural representations.
\end{abstract}

\section{Introduction}
\label{sec:introduction}

The brain learns remarkably rich representations from sensory experience. Explaining how such powerful learning can emerge from largely unsupervised, biologically plausible local computations is a central challenge in neuroscience. The Bayesian Brain family of theories offers one influential account, with Predictive Coding (PC) as a special case proposing that learning is driven by hierarchical, locally computed prediction errors. This account has been highly successful in explaining a wide range of perceptual and cognitive phenomena \citep{rao1999predictive, summerfield2008neural, hohwy2008predictive, garrido2009mismatch, feldman2010attention, auksztulewicz2016repetition, walsh2020evaluating, feuerriegel2021evaluating, aizenbud2025neural}.

While a complete computational formalization of PC remains a matter of debate \citep{shipp2016neural, millidge2024predictive}, a general consensus can be identified regarding requirements a given system must fulfil to qualify as implementing of PC (from here on “PC-compliant”). \textbf{Requirement 1}: the system is organized hierarchically, with each level repeating the same computational task at a different level of abstraction \citep{sprevak2024predictive, walsh2020evaluating, millidge2021predictive}. \textbf{Requirement 2}: part of the system performs the role of a generative model, most commonly implemented as a top-down process that predicts incoming bottom-up information \citep{spratling2017review, sprevak2024predictive, walsh2020evaluating, millidge2021predictive}. \textbf{Requirement 3}: learning is driven based on the mismatch (prediction error) between this top-down prediction and the bottom-up input \citep{spratling2017review, sprevak2024predictive, walsh2020evaluating, millidge2021predictive}. \textbf{Requirement 4}:, the predictions and errors are represented in functionally, anatomically, or structurally distinct neuronal populations \citep{spratling2017review, sprevak2024predictive, walsh2020evaluating, millidge2021predictive}. \textbf{Requirement 5}: learning is implemented through spatially local mechanisms that update synaptic connection weights \citep{spratling2017review, sprevak2024predictive, millidge2021predictive}. \textbf{Requirement 6}: the predictions created are of probabilistic, or bayesian nature \citep{spratling2017review, sprevak2024predictive, millidge2021predictive}.

Beyond these mechanistic requirements, successful learning is generally expected to transform high-dimensional sensory inputs into compact, abstract representations that capture underlying structure \citep{edelman1997learning, gibson1955perceptual, wolff2019information}. Accordingly, a successful PC-compliant system should implement the above computational requirements to learn such abstract representations of sensory information, as is also observed in the brain \citep{edelman1997learning, gibson1955perceptual, wolff2019information}.

In recent years, an alternative approach for computational neuroscience has gained traction: Deep Neural Network (DNN) models are used to formalize neuroscientific hypotheses through their architectures, objectives, training data, and learning rules, and the resulting activations and behavior are compared to those observed in the brain (an approach referred to as \textit{neuroconnectionism}) \citep{doerig2023neuroconnectionist, golan2023deep, cichy2019deep}. Through this approach, established DNNs originally created for engineering goals and only loosely inspired by biology  \citep{he2016deep, liu2022convnet, dosovitskiy2020image} have proven to learn rich and generally applicable representations which provide some of the strongest predictors of brain activity in vision \citep{cichy2016comparison, kietzmann2019recurrence, gucclu2015deep, khaligh2014deep, doerig2025high}, audition \citep{kell2018task, saddler2021deep}, language \citep{caucheteux2022brains, schrimpf2021neural}, and more \citep{schrimpf2018brain, conwell2022can}.

The computationally well-defined and empirically supported PC account of perception and learning in neural systems should be a natural fit for neuroconnectionism. If PC captures important brain computations, then integrating the relevant computational characteristics into PC-compliant DNNs should lead to powerful brain models. However, a striking paradox has persisted: PC-compliant computational models have so far drastically failed to produce rich brain-like representations on par with those of established DNN  \citep{pinchetti2025benchmarking, biscione2023mixed}.

This disadvantage of PC-compliant DNNs is not for lack of effort. Numerous projects have set out to instantiate the PC theoretical framework in computational models \citep{lotter2016deep, jiang2024dynamic, whittington2017approximation, wen2018deep, millidge2024predictive, innocenti2026mu, kerjan2026training}. So far, none of these approaches have succeeded in strictly adhering to the principles of PC while producing representations that compete with those of other established DNN architectures in terms of brain alignment and ability to perform downstream tasks \citep{pinchetti2025benchmarking, biscione2023mixed}. A prominent example is PredNet \citep{lotter2016deep}, a highly influential system that sparked interest in PC-inspired DNNs. It successfully implemented several of our stated requirements (1, 2, 3, and 4) in an end-to-end trainable neural network and performed only slightly below other unsupervised DNNs of the time on next frame prediction (38.9\% peak-signal-to-toise-ratio on \citep{ionescu2013human3} compared to 42.1\% for \citep{finn2016unsupervised}; see \citep{lotter2016deep}). However, it has substantially lower brain-alignment than other DNN architectures \citep{zhuang2021unsupervised}, lacks local error minimization (requirement 5) \citep{rane2020prednet}, in its original implementation does not produce bayesian or probabilistic predictions (requirement 6) \citep{lotter2016deep}, and fails to replicate experimental hallmarks of PC \citep{biscione2023mixed, brands2024deep, kirubeswaran2023inconsistent}.

This limitation of existing PC-inspired models raises two major issues. First it restricts the applicability of PC-inspired computational models to simple settings rather than the more complex naturalistic-input settings where modern DNNs thrive \citep{deco2001predictive, spratling2010predictive, auksztulewicz2015attentional, pinotsis2017linking, pouget2013probabilistic, rane2020prednet}, Second, it leaves PC without a compelling computational demonstration that its proposed mechanism is sufficient for learning rich, brain-aligned representations. This leads to an apparent paradox: how can it be that a theory of brain computations as developed, influential, and empirically substantiated as PC does not lead to competitive models to learn rich and brain-aligned computations?

Here, we show that this paradox can be overcome. We introduce Predictive Brain Networks (PBNs), a general framework for constructing fully PC-compliant models that combine principles of PC with the representational power of modern DNNs. First,  we show that PBNs far surpass previous state-of-the-art PC-inspired DNNs by achieving representational alignment with the human visual hierarchy drastically better than PredNet \citep{lotter2016deep} and on par or better than ResNet-50 \citep{he2016deep} and ConvNeXt \citep{liu2022convnet}, closing the decade-long gap in brain alignment left open by PC-inspired models. Second, the brain alignment of PBNs reflects specifically predictive computations: top-down predictive decoder representations explain neural variance beyond bottom-up encoders, with this contribution increasing as learning becomes more local. Third, PBNs replicate key empirical and theoretical signatures of PC: top-down predictions observed in macaque face processing \citep{schwiedrzik2017high}. Additionally, we show that the representations learnt by PBNs are sufficiently rich to successfully perform downstream tasks and that PBNs are capable of modeling uncertainty using \citep{gal2016dropout}.

\begin{figure}[htbp]
  \centering
  \includegraphics[width=1.\linewidth]{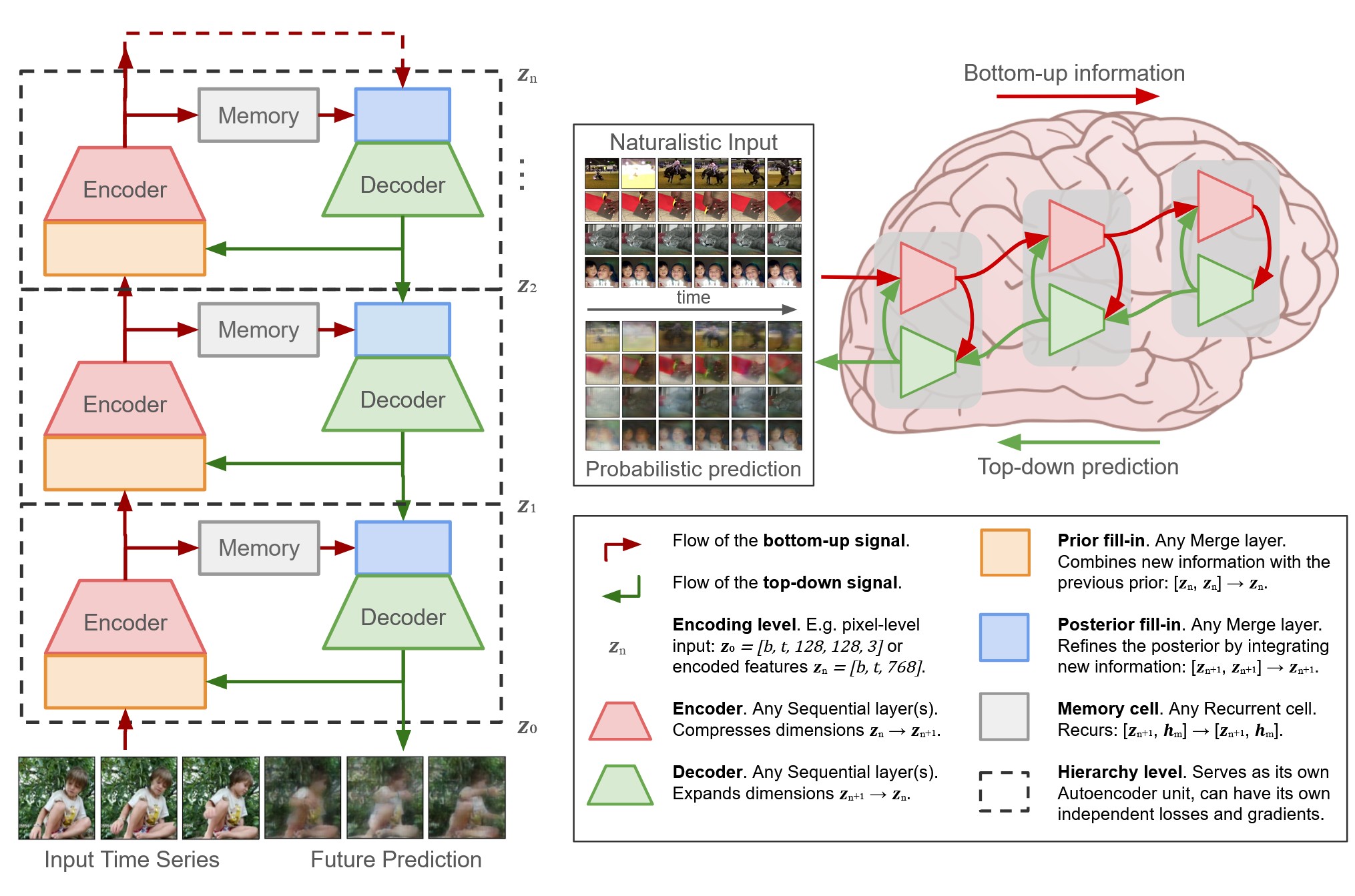}
  \caption{\textbf{General overview of the architecture.} \textbf{Left:} Architecture of the Predictive Brain Networks (PBN), illustrating the functional flow of information in a deep-learning framework based on a hierarchy of recurrent autoencoders with skip-connections while highlighting its correspondence to the PC architecture.  \textbf{Bottom right:} Legend and description of the network architecture elements shown in the left panel. \textbf{Top right:} Schematic of learning according to PC. Learning of abstract categories is achieved through hierarchically repeating elements that locally encode stimuli by minimizing errors between bottom-up inputs and probabilistic generative top-down predictions.}
  \label{fig:general-overview}
\end{figure}

\section{Results}

\subsection{General architecture}

To construct PC-compliant DNNs that satisfy requirements 1-6 listed earlier, we developed a flexible framework that we refer to as predictive brain networks (PBNs). PBNs can be described as a hierarchy of recurrent autoencoders. At each hierarchical level, an encoder maps inputs into latent representations and a decoder uses these representations to predict future inputs (for an overview, see Figure \ref{fig:general-overview}). These autoencoders are stacked such that each higher-level unit encodes the latent representation of the level below, while skip connections allow information exchange across the hierarchy. This hierarchical organization enables each level to be trained using a local loss that compares its prediction with its input, while restricting gradient flow between levels can be used to enforce local learning while still allowing the formation of high-level abstract representations (for details, see \hyperref[sec:model-architecture]{Methods}, and Figure \ref{fig:general-overview}).

This architecture addresses the key computational requirements of PC. Its hierarchical organization consists of repeating computational blocks (requirement 1) each implementing a generative model (requirement 2). Learning is driven by prediction errors between bottom-up inputs and top-down predictions at each hierarchical level (given a PC-compliant loss; requirement 3), with these signals represented in distinct populations (requirement 4). Gradients used for learning can be restricted to a given hierarchical level to enforce strictly local learning (requirement 5). The probabilistic nature of predictions (requirement 6) can be addressed by combining the loss with dropout and weight decay to implement Monte Carlo dropout as an approximate Bayesian inference method \citep{gal2016dropout} (see \hyperref[sec:methods-uncertainty-quantification]{Methods}). Together, these components enable the model to learn rich, brain-aligned representations across hierarchical levels.

The PBN framework is designed to maximize flexibility: it can be used for different data domains and network architectures, and both loss functions and gradient flow can be configured flexibly. This enables a broad range of models to be implemented within a common framework, provided they retain the core structure of a hierarchical, autoregressive autoencoder. Our aim is to facilitate the development and comparison of alternative PBN variants, allowing researchers to systematically implement and test competing neuroscientific hypotheses about PC, learning, and perception.

\subsection{Model variants}

We trained PBNs on naturalistic video sequences \citep{monfort2019moments, gifford2024algonauts, dave2020tao} and compared them with established vision models used in neuroconnectionist research. We focused on vision because it represents the largest and most established modality, with widely used models and datasets. To investigate how PC-specific learning constraints shape learned representations, we constructed six PBN variants that share the same underlying architecture while varying PC-related components (see Table \ref{tab:overview-of-pbn-variants-the-first-4-variants-pr}, Figure \ref{fig:model_variant_architecture_visualization}). Our principal variant, \textbf{PBN local gradients}, satisfies all six PC requirements outlined above, while producing rich, brain-aligned representations. It uses a Contrastive Predictive Coding (CPC) \cite{oord2018representation} loss at each hierarchical level, and purely local learning (by using blocked gradient flow between hierarchy levels).

As our first group of control conditions, we constructed three model variants that contrast our PBN local gradients variant with less PC-compliant variants. \textbf{PBN untrained} shares the same architecture but receives no training, providing an architectural baseline for representations in the absence of learning. \textbf{PBN backprop }uses standard backpropagation \citep{rumelhart1986learning}, with a CPC loss at the highest level and a reconstruction loss at the lowest level (equivalent to a U-Net autoencoder \citep{ronneberger2015u}, with a contrastive loss at the bottleneck and a reconstructive objective), providing a baseline for conventional end-to-end learning with a single high-level contrastive objective. \textbf{PBN backprop + local loss} applies local CPC losses at each hierarchy level but allows gradients to flow between levels. Comparing this variant with PBN local gradients therefore tests the impact of strictly local learning when layer-specific predictive objectives are already in place.

We additionally considered two variants that satisfy all six PC requirements but adopt implementation choices more closely aligned with classic PC \citep{rao1999predictive}. \textbf{PBN local gradients + subtract} extends the principal condition by explicitly subtracting the top-down prior from the bottom-up input, as proposed in classic PC formulations \citep{rao1999predictive}. Finally, \textbf{PBN classic PC }extends this subtraction based condition by also using a non-contrastive augmentation-free loss. While all other PBN variants combine light data augmentations (intended to approximate aspects of natural variability in visual input, see \hyperref[sec:methods-model-variants]{Methods}) with a contrastive CPC objective \citep{oord2018representation} that pulls together valid predictions and future inputs and pushes apart invalid predictions and future inputs, PBN classic PC uses invariance-based mean-squared-error (MSE) loss without augmentations, allowing us to investigate the effect of contrastive predictive learning compared to invariance-based predictive learning.

To compare these PBN variants with established DNN models, we included \textbf{ResNet-50} \citep{he2016deep} and \textbf{ConvNeXt-tiny} \citep{liu2022convnet}, two standard vision models trained under supervised conditions on ImageNet and selected among the top-performing models on Brain-Score \citep{schrimpf2018brain}, a standard benchmark for computer models of human visual neurophysiological data. We additionally included \textbf{PredNet }\citep{lotter2016deep}, the currently most cited PC network (see \hyperref[sec:introduction]{Introduction}). PredNet was matched to the PBN models in parameter count to within 3\% and trained on the same data as the PBNs, using the hyperparameter settings of the original publication.

All trained model variants were trained on \citep{monfort2019moments, gifford2024algonauts, dave2020tao} using an early stopping procedure (see \hyperref[sec:methods-training]{Methods}) and converged in less than 350 epochs, with the subtraction conditions (PBN local gradients + subtract and PBN classic PC) converging notably earlier than other variants (before epoch 100).

\begin{table}[bt]
  \centering
  \caption{Overview of PBN variants. The first 4 variants progressively introduce local predictive objectives and local learning, culminating in PBN local gradients (as our main fully PC-compliant condition). The last 2 variants retain all six PC requirements while adopting increasingly classic PC implementations through explicit prior subtraction and, in PBN classic PC, invariance-based learning without augmentations.}
  \label{tab:overview-of-pbn-variants-the-first-4-variants-pr}

  \begin{tabular}{llllll}
    \hline
    Name & Loss & \makecell{Local\\Loss} & \makecell{Local\\Gradients} & Subtract & \makecell{Augmentation-\\free} \\
    \hline
    PBN untrained
      & \cellcolor{noColor}No
      & \cellcolor{noColor}No
      & \cellcolor{noColor}No
      & \cellcolor{noColor}No
      & \cellcolor{noColor}No \\

    PBN backprop
      & \cellcolor{cpcColor}CPC
      & \cellcolor{noColor}No
      & \cellcolor{noColor}No
      & \cellcolor{noColor}No
      & \cellcolor{noColor}No \\

    PBN backprop + local loss
      & \cellcolor{cpcColor}CPC
      & \cellcolor{yesColor}Yes
      & \cellcolor{noColor}No
      & \cellcolor{noColor}No
      & \cellcolor{noColor}No \\

    \textbf{PBN local gradients}
      & \cellcolor{cpcColor}\textbf{CPC}
      & \cellcolor{yesColor}\textbf{Yes}
      & \cellcolor{yesColor}\textbf{Yes}
      & \cellcolor{noColor}\textbf{No}
      & \cellcolor{noColor}\textbf{No} \\

    PBN local gradients + subtract
      & \cellcolor{cpcColor}CPC
      & \cellcolor{yesColor}Yes
      & \cellcolor{yesColor}Yes
      & \cellcolor{yesColor}Yes
      & \cellcolor{noColor}No \\

    PBN classic PC
      & \cellcolor{mseColor}MSE
      & \cellcolor{yesColor}Yes
      & \cellcolor{yesColor}Yes
      & \cellcolor{yesColor}Yes
      & \cellcolor{yesColor}Yes \\
    \hline
  \end{tabular}
\end{table}

\subsection{Representational similarity to the brain}
\label{sec:results-rsa}

\begin{figure}[htbp]
  \centering
  \includegraphics[width=1.\linewidth]{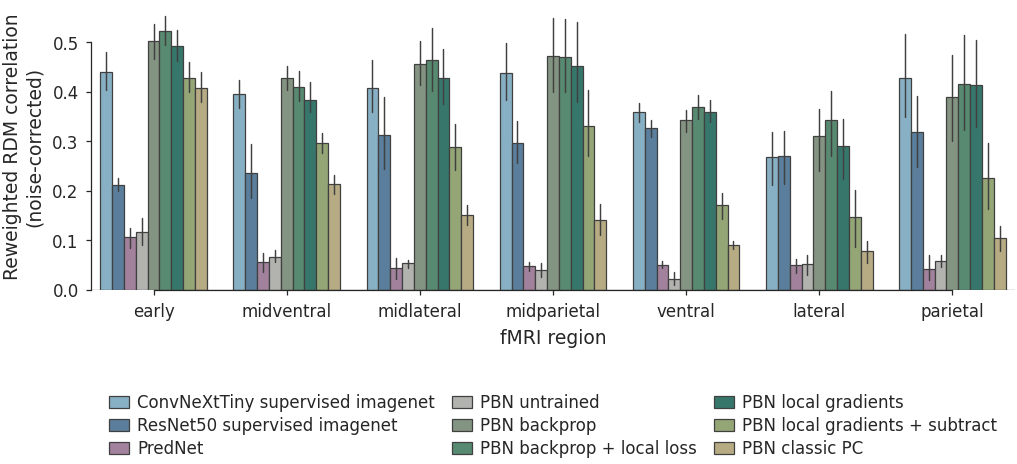}
  \caption{\textbf{Representational similarity between model representations and human visual cortex.} Similarity was quantified using noise-ceiling-corrected Spearman correlations between reweighted network representations, and mean fMRI responses to the same images from the Natural Scenes Dataset \citep{allen2022massive}, across seven visual regions. Colours indicate established vision DNNs (ResNet-50 and ConvNeXt-tiny), PredNet, and PBN variants PBN untrained, PBN backprop, PBN backprop + local loss, PBN local gradients (the principal condition), PBN local gradients + subtract and PBN classic PC. PBN local gradients matched or exceeded established DNNs, significantly outperforming them in seven of 14 comparisons, with no comparisons significantly favouring established DNNs. The less PC-compliant PBN controls also showed strong similarity, whereas prior-subtraction variants showed lower similarity. All trained PBN variants outperformed PredNet across all seven regions, while the untrained PBN did not differ significantly from PredNet.}
  \label{fig:representational-similarity-between-model-repres}
\end{figure}

\begin{figure}[htbp]
  \centering
  \includegraphics[width=1.\linewidth]{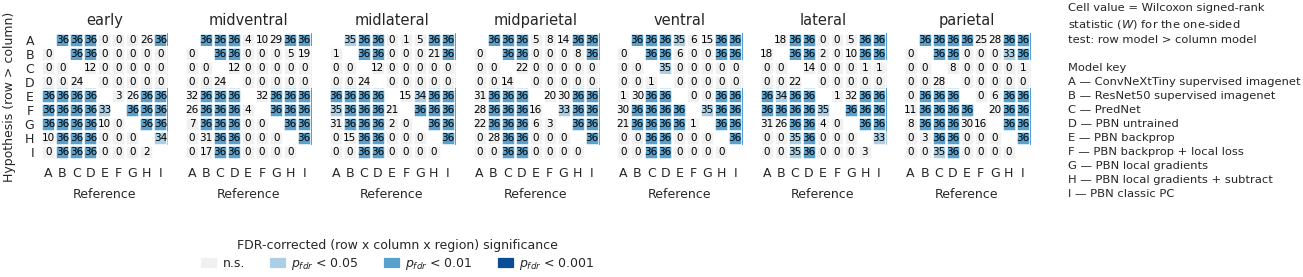}
  \caption{\textbf{Post-Hoc pairwise comparisons between all models reported in \hyperref[sec:results-rsa]{Results/Respresentational similarity to the brain}.} Models are coded as: A, ConvNeXt Tiny supervised imagenet; B, ResNet50 supervised imagenet; C, PredNet; D, PBN untrained; E, PBN backprop; F, PBN backprop + local loss; G, PBN backprop + local gradients; H, PBN local gradients + subtract; I, PBN classic PC. Each element depicts a single-sided Wilcoxon test, testing if row > column. Numbers show the test statistic Wilcoxon’s W. Background colors show FDR-corrected (over conditions x columns x regions) significance levels.}
  \label{fig:rsa_reweighted_pairwise}
\end{figure}

We first evaluated how closely PBN representations matched human visual representations and compared them with established vision DNNs. We compared model representations with human responses to the 515 images shared across participants in the Natural Scenes Dataset (NSD) \citep{allen2022massive}, a large-scale fMRI dataset of human vision, using reweighted \citep{storrs2021diverse} and noise-ceiling-corrected representational similarity analysis (RSA) \citep{kriegeskorte2008representational} across seven visual regions (see Figure \ref{fig:representational-similarity-between-model-repres}, for pairwise comparisons see Figure \ref{fig:rsa_reweighted_pairwise}). Our principal model, PBN local gradients, significantly outperformed established DNNs in 7 of 14 region-by-model comparisons (Wilcoxon signed-rank tests, \(p_{\mathrm{FDR}} < 0.01\) ). The two less PC-compliant controls, PBN backprop and PBN backprop + local loss, also showed strong representational similarity to human visual responses. Across their 28 comparisons with established DNNs, these variants significantly outperformed the DNNs in 16 (Wilcoxon signed-rank tests, \(p_{\mathrm{FDR}} < 0.05\) ) comparisons and were significantly outperformed in two (Wilcoxon signed-rank tests, \(p_{\mathrm{FDR}} < 0.05\) ). In contrast, the two variants implementing stricter classic PC assumptions, PBN local gradients + subtract and PBN classic PC, showed lower representational similarity to established DNNs. Across their 28 comparisons, established DNNs significantly outperformed these variants in 21 (Wilcoxon signed-rank tests, \(p_{\mathrm{FDR}} < 0.05\) ) comparisons, whereas the subtractive PBN variants significantly outperformed the established DNNs in 2  (Wilcoxon signed-rank tests, \(p_{\mathrm{FDR}} < 0.01\) ), with subtractive PBN variants performing well in early visual cortex, but drastically dropping in higher areas. All five trained PBN variants significantly outperformed PredNet across all seven visual regions, comprising all 35 model-by-region comparisons (Wilcoxon signed-rank tests, \(p_{\mathrm{FDR}} < 0.05\) ). By contrast, the untrained PBN baseline did not differ significantly from PredNet in any of the seven areas (see Figure \ref{fig:rsa_reweighted_pairwise}). Together, these results show that the PBN architecture can learn representations whose similarity to human visual representations matches or exceeds that of established vision DNNs. This result holds for the principal PC-compliant condition and the two less PC-compliant controls. The two prior-subtraction variants, while still substantially outperforming PredNet across all seven areas, show lower representational similarity than established DNNs, which suggests that explicit prior subtraction may limit brain-representational similarity. As a confirmatory analysis, we additionally implemented an alternative approach that instead of reweighted RSA uses the model layer representation with the highest correlation to the corresponding brain area. This analysis is reported in Figure \ref{fig:rsa_maximum} and yielded qualitatively similar results.

\subsection{PBN decoders capture unique information}

\begin{figure}[htbp]
  \centering
  \includegraphics[width=1.\linewidth]{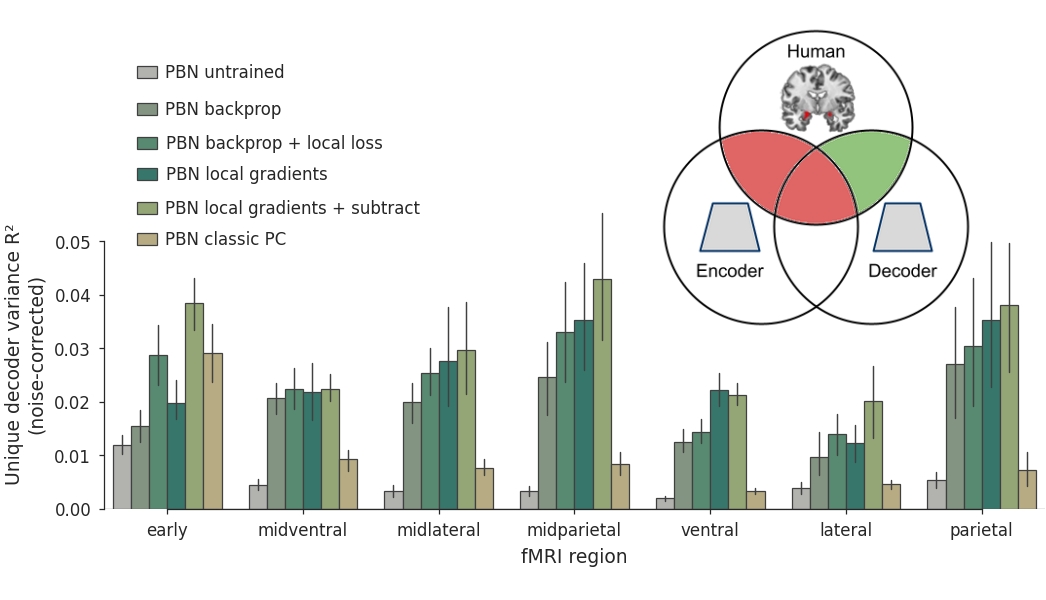}
  \caption{\textbf{Unique variance explained by the decoder in brain data, excluding encoder-explained variance.} \textbf{Main:} Noise-ceiling-corrected unique variance (R²) explained by the PBN decoder in human fMRI responses, after factoring out variance explained by the encoder, for naturalistic images from \citep{allen2022massive} NSD 515 shared images. Bars show the mean across subjects for each model variant and brain region; error bars indicate 95\% confidence intervals. \textbf{Top right:} Venn diagram illustrating the partition of unique decoder-explained variance (green) and the variance partitions factored out as encoder-explained variance (red).}
  \label{fig:figure-3-unique-variance-explained-by-the-decode}
\end{figure}

\begin{figure}[htbp]
  \centering
  \includegraphics[width=1.\linewidth]{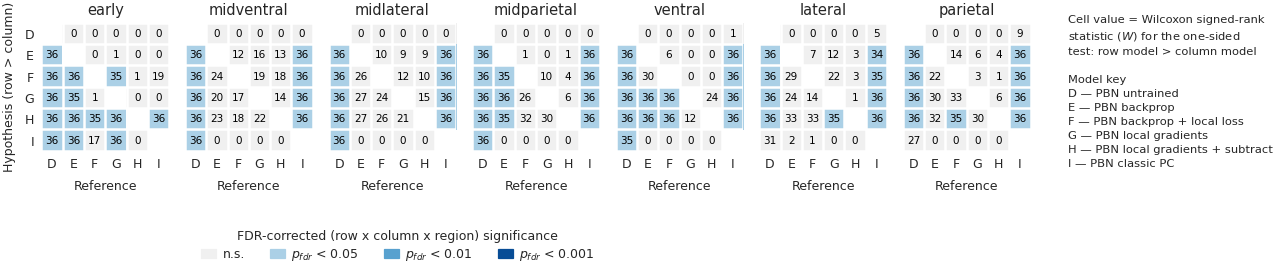}
  \caption{\textbf{Pairwise post-hoc comparisons of unique decoder-explained variance.} Noise-ceiling-corrected unique variance (R²) explained by the decoder in each brain region, after factoring out variance explained by the encoder, compared between models. Models are labelled as follows: D, PBN untrained; E, PBN backprop; F, PBN backprop + local loss; G, PBN local gradients; H, PBN local gradients + subtract; and I, PBN classic PC. Each cell shows a one-sided Wilcoxon signed-rank test, paired by subject, of whether the row model explains more unique decoder variance than the column model. Numbers indicate the Wilcoxon test statistic, W, and background colours indicate significance levels after false discovery rate (FDR) correction across model pairs and brain regions.}
  \label{fig:decoder_unique_var_pairwise}
\end{figure}

Next, we tested whether the decoder of a PBN explains variance in human visual responses that the encoder does not. This distinction is important because the decoder captures additional brain-response variance through the predictive, recurrent and top-down processing central to PC, which conventional feedforward models cannot capture. We quantified decoder-unique variance for each model as the difference between the variance explained jointly by all encoder and decoder layers and that explained by all encoder layers alone, \(R^2(\text{brain} \mid \text{all encoder layers} \cup \text{all decoder layers}) - R^2(\text{brain} \mid \text{all encoder layers})\). We then compared all models (see Figure \ref{fig:figure-3-unique-variance-explained-by-the-decode}) using pairwise comparisons (for details see Figure \ref{fig:decoder_unique_var_pairwise} ).

All CPC-based PBN variants explained significantly more unique decoder variance than the PBN untrained baseline in all 28 comparisons (Wilcoxon signed-rank tests, \(p_{\mathrm{FDR}} < 0.013\) ), indicating that decoder-unique variance reflects learned features rather than architecture alone (see Figure \ref{fig:figure-3-unique-variance-explained-by-the-decode}). Our main model, PBN local gradients, matched or exceeded the less constrained PBN backprop variant across all regions, explaining significantly more unique decoder variance in the early, midparietal and ventral regions (Wilcoxon signed-rank tests, \(p_{\mathrm{FDR}} < 0.022\) ). PBN backprop did not exceed PBN local gradients in any region. This pattern is consistent with an increase in unique decoder information as PC constraints are strengthened, a trend also observed across the other CPC-based PBN variants (see Figure \ref{fig:decoder_unique_var_pairwise}). The non-CPC-based PBN classic PC variant was the exception: although it did not reach the unique decoder variance of the CPC-based models, it still explained significantly more unique variance than the PBN untrained baseline in the early, midventral, midlateral, midparietal and ventral regions (Wilcoxon signed-rank tests, \(p_{\mathrm{FDR}} < 0.022\) ), but not in the lateral or parietal regions. Together, these results indicate that CPC-based PBN variants learn top-down decoder representations that explain brain variance beyond their bottom-up encoder representations, consistent with Bayesian brain theories. The contribution is strongest in the more PC-compliant variants and weaker in the PBN classic PC model. Importantly, PBN local gradients not only matches established DNNs in brain alignment but also captures unique brain-related variance through its decoder, highlighting the value of its PC-compliant architecture.

\subsection{PBNs outperform prior PC-inspired DNNs on downstream classification tasks}

An important expectation of a learning framework (including PC) is that it learns rich, compact representations that are broadly useful \citep{edelman1997learning, gibson1955perceptual, wolff2019information, spratling2017review, sprevak2024predictive, millidge2021predictive}. To assess this, we evaluated the downstream performance of the learned representations of all compared models on a separate action-recognition task. We evaluated all models on the test set of the Moments in Time \citep{monfort2019moments} and the UCF-101 \citep{soomro2012ucf101} action-recognition datasets. For each model, we extracted activations from the final encoder layer and concatenated the activations across all timesteps; for PBN, we used the encoder representation, whereas for PredNet we used all recurrent states at the corresponding layer. We applied both a K-nearest-neighbour readout (KNN-1) and a linear readout, calculating classification accuracy for each action category (for details, see \hyperref[sec:methods-readout-performance]{Methods}). The PBN backprop, PBN backprop + local loss, and our PC-compliant PBN local gradient variants approached ResNet50 and ConvNeXT performance, and consistently outperformed PredNet across both datasets and both readout metrics. This demonstrates that the PBN architecture substantially advances the ability of PC-inspired models to learn representations for abstract visual recognition (see Table \ref{tab:downstream-performance-accuracy-on-the-ucf-101-d}).

\begin{table}[bt]
  \centering
  \caption{Downstream performance accuracy on the UCF-101 dataset and the Moments in Time test set. All cells show mean accuracy ± standard deviation across 5-fold cross-validation splits. Best performing model shown in bold. Established networks perform best, while PBN backprop, PBN backprop + local loss, and PBN backprop local gradients consistently outperform PredNet and perform closer to the performance of the established models.}
  \label{tab:downstream-performance-accuracy-on-the-ucf-101-d}
  \setlength{\tabcolsep}{4pt}
  \begin{tabular}{@{}lcccc@{}}
    \hline
    \textbf{Model} & \multicolumn{2}{c}{\textbf{UCF-101}} & \multicolumn{2}{c}{\textbf{Moments in Time}} \\
    \cline{2-3}\cline{4-5}
     & kNN-1 & Linear readout & kNN-1 & Linear readout \\
    \hline
    ConvNeXtTiny (ImageNet) & 0.881 ± 0.009 & \textbf{0.464 ± 0.009} & \textbf{0.059 ± 0.001} & \textbf{0.102 ± 0.003} \\
    ResNet50 (ImageNet) & \textbf{0.921 ± 0.005} & 0.337 ± 0.010 & 0.034 ± 0.001 & 0.064 ± 0.002 \\
    PredNet & 0.773 ± 0.011 & 0.071 ± 0.006 & 0.006 ± 0.001 & 0.003 ± 0.000 \\
    PBN untrained & 0.634 ± 0.011 & 0.039 ± 0.007 & 0.006 ± 0.001 & 0.003 ± 0.000 \\
    PBN backprop & 0.819 ± 0.011 & 0.302 ± 0.016 & 0.018 ± 0.001 & 0.043 ± 0.001 \\
    PBN backprop + local loss & 0.849 ± 0.013 & 0.309 ± 0.014 & 0.021 ± 0.001 & 0.044 ± 0.002 \\
    PBN local gradients & 0.828 ± 0.015 & 0.270 ± 0.014 & 0.016 ± 0.002 & 0.038 ± 0.002 \\
    PBN local gradients + subtract & 0.338 ± 0.014 & 0.026 ± 0.004 & 0.006 ± 0.001 & 0.006 ± 0.001 \\
    PBN classic PC & 0.854 ± 0.005 & 0.237 ± 0.018 & 0.010 ± 0.001 & 0.019 ± 0.002 \\
    \hline
  \end{tabular}
\end{table}

\subsection{Higher-level predictions reshape expectations in lower layers in brains and PBNs}

\begin{figure}[htbp]
  \centering
  \includegraphics[width=1.\linewidth]{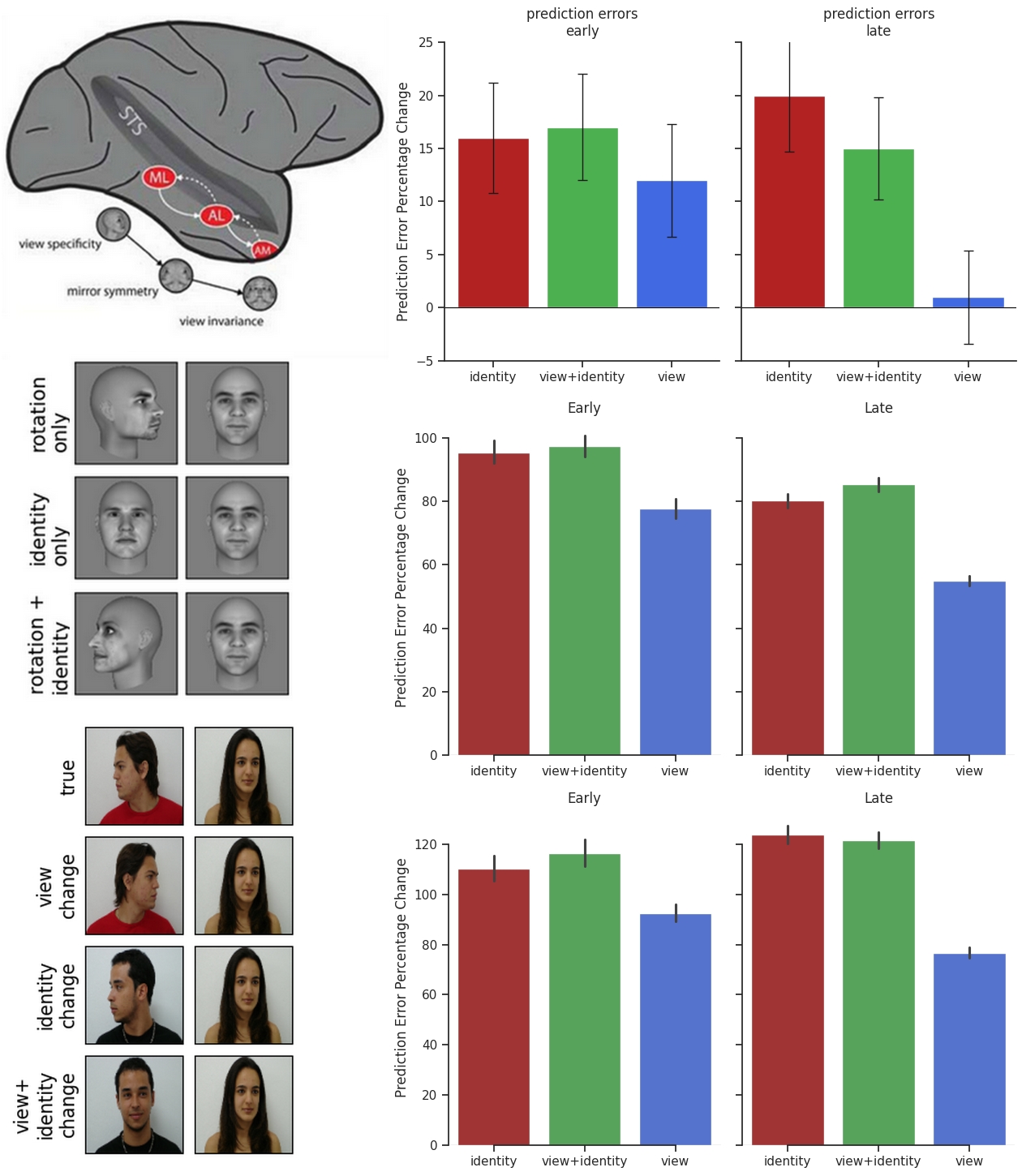}
  \caption{\textbf{Replication of \cite{schwiedrzik2017high}}. \textbf{Top left:} Schematic adapted from the original study, illustrating information exchange in the macaque cortex. \textbf{Center left:} Stimulus categories used in the original study. \textbf{Bottom left:} Stimulus categories used to train the PBN, based on \citep{thomaz2010new}, which reproduces the stimulus categories of the original paradigm with more samples and more naturalistic images. \textbf{Top right:} Main results from the original study, showing the interaction between early versus late time steps and identity versus view violations. \textbf{Center right:} Our replication results in layer 1. Prediction errors consistently decrease after the first step, indicating rapid adaptation, while the view-violation condition shows a significantly greater decay than the identity-violation condition (interaction: t = 4.841, p < 1e-6). \textbf{Bottom right:} Our replication results in layer 2, showing the same pattern with a similar interaction (t = 13.900, p < 1e-37).}
  \label{fig:replication_freiwald}
\end{figure}

Having established that the PBN local gradients model learns rich, brain-aligned representations while remaining fully PC-compliant, we now tested whether this PBN variant can model neuroscientific hallmarks of PC that are not captured by feedforward DNNs (Figure \ref{fig:replication_freiwald}). PC entails that predictions from higher-level representations can shape the information in lower-level areas. In face processing, lower-level areas are more sensitive to face view, whereas higher-level areas represent identity across different views. \cite{schwiedrzik2017high} tested in macaque monkeys whether learned predictions allow these higher-level representations to shape lower-level prediction errors. After macaques learned sequences of face pairs, the researchers changed the preceding face to violate the learned association in identity (a different person), view (the same person from a different angle), or both, while keeping the subsequent face unchanged. They compared early neural responses with late responses to test whether prediction errors changed over time as recurrent feedback could influence processing. In the lower view-specific area, view violations elicited prediction errors early on, but these errors diminished in the late response window, whereas identity-violation errors persisted. This pattern indicates that lower-level prediction errors increasingly reflect the identity-specific, view-invariant properties of higher-level representations, consistent with top-down predictions shaping lower-level processing.

To test whether PBNs reproduce this temporal signature, we fine-tuned the PBN local gradients variant to predict face sequences from \citep{thomaz2010new}, reproducing the experimental setup from \citep{schwiedrzik2017high}. At test, we introduced the same identity, view, and combined violations by changing the preceding face while keeping the subsequent image constant (see \hyperref[sec:methods-replication]{Methods}). We analyzed prediction errors in the first two PBN layers, hypothesized to correspond to the low-level stages examined in the macaque study, and tested the interaction between violation type and early versus late window using paired difference-of-differences t-tests \citep{student1908probable}. This interaction was highly significant in both layers (Layer 1: \(t = 4.841,\ p < 10^{-6}\); Layer 2: \(t = 13.900,\ p < 10^{-37}\) ). Like in macaques, prediction errors to view violations declined from early to late responses, whereas identity-violation errors were more sustained. PBNs thus reproduce the temporal signature observed in macaques, consistent with higher-level predictions shaping lower-level prediction errors.

\FloatBarrier

\subsection{PBNs can quantify uncertainty using Dropout Monte Carlo}

Another hallmark of Bayesian brain and PC theories, is to provide a way to represent uncertainty about its predictions. One common approach in probabilistic PC models is variational inference, in which uncertainty is represented by an approximate posterior distribution over latent variables. For example, variational autoencoders parameterize the posterior over each latent variable using a predefined distribution, typically a unimodal Gaussian \citep{kingma2013auto}. This makes uncertainty explicit, but also constrains the form that the latent posterior can take. An alternative is to approximate Bayesian inference through uncertainty over the network's weights rather than by explicitly specifying a parametric distribution over latent representations. \cite{gal2016dropout} mathematically proved that any network trained with dropout and weight decay under a standard loss can be interpreted as performing approximate variational inference over network weights. In line with this approach, we estimate prediction uncertainty in PBNs using Monte Carlo dropout \citep{gal2016dropout}. For a given input, we perform multiple forward passes with dropout enabled, producing a set of stochastic predictions. The distribution of these predictions provides an estimate of the model's predictive uncertainty \citep{gal2016dropout}. This procedure has a natural analogue in sampling-based theories of neural computation, which propose that stochastic neural activity represents samples from an underlying posterior distribution over possible interpretations of the sensory input \citep{barber2003neural, ma2006bayesian, orban2016neural}, and can therefore be seen as a more natural analogy than strictly parametrizing variational methods.

\begin{figure}[htbp]
  \centering
  \includegraphics[width=1.\linewidth]{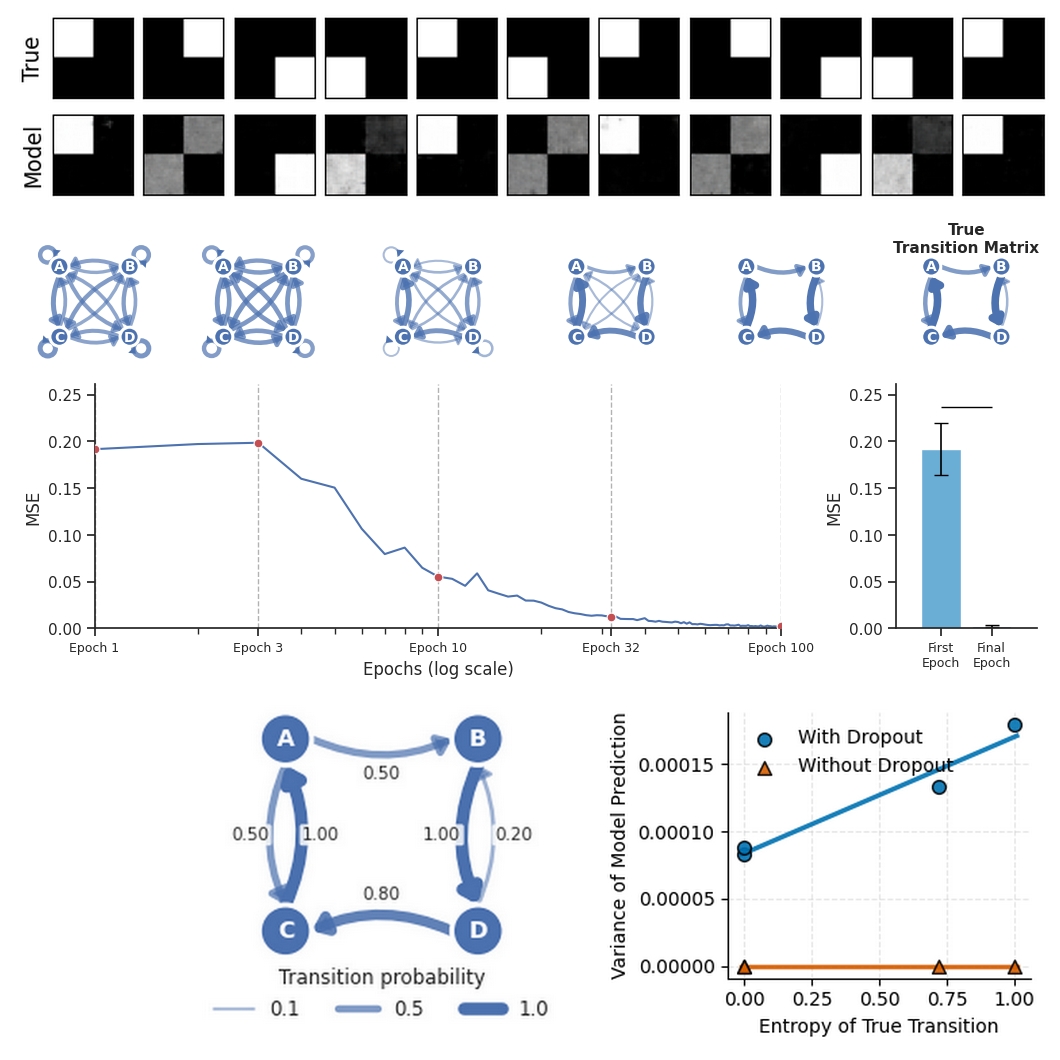}
  \caption{\textbf{Markov chain experiment for uncertainty quantification. Top:} Markov chain stimulus images presented to the model and the corresponding model predictions after 100 epochs of fine-tuning. \textbf{Center left:} Mean squared error (MSE) between the true Markov transition matrix and the transition matrix inferred from model outputs over 100 (logarithmically presented) epochs of training. At logarithmically spaced epochs indicated via dashed lines, model-inferred Markov chains are visualized above. \textbf{Center right:} MSE between the true and inferred transition matrices after epoch 1 and epoch 100, showing a clear reduction in error (Wilcoxon W = 135, p < 1e-4). \textbf{Bottom left:} True Markov chain with the transition probability specified for each corner. \textbf{Bottom right:} Variance under Dropout Monte Carlo represents true transition entropy. Mean variance for each transition state plotted against the overall entropy of that transition state with and without Dropout Monte Carlo. Lines indicate linear fits for each condition. While the same model with Dropout turned off logically produces no variance for different entropy transitions, the variance under Dropout Monte Carlo scales with transition entropy, and can therefore be used to quantify model uncertainty as established by \citep{gal2016dropout}.}
  \label{fig:markov-chain-experiment}
\end{figure}

To demonstrate PBN capability of uncertainty quantification, we devised a visual Markov-chain experiment in which four possible states were represented by white corners in an otherwise black input frame, with transitions between states governed by a predefined Markov transition matrix (see Figure \ref{fig:markov-chain-experiment}, top). For example, when the top-left corner (A) was active in the preceding frame, the probabilities of the top-left (A), top-right (B), bottom-left (C), and bottom-right (D) corners being active in the next frame were 0\%, 50\%, 50\%, and 0\%, respectively (the complete transition structure is shown in Figure \ref{fig:markov-chain-experiment}, bottom left). We fine-tuned the PBN local gradients model on these sequences for 100 epochs (for details, see \hyperref[sec:methods-uncertainty-quantification]{Methods}) and evaluated it on a separate test set. After positively confirming that the model is capable of doing next-frame prediction (reported in Figure \ref{fig:figure-a5-future-prediction-on-the-markov-datase}), we investigated whether the model is capable of learning the inherent rules of a markov chain and of correctly representing its inherent uncertainty. To do so, we first evaluated how accurately the model recovered the transition structure by measuring the predicted activity for each of the four possible states. During training, the predicted transition structure converged from an initially near-random configuration towards the transition matrix used to generate the stimuli, resulting in a significant difference between the first and final epochs (Wilcoxon \(W = 135\), \(p < 10^{-4}\); see Figure \ref{fig:markov-chain-experiment}, center). We next tested whether the stochasticity introduced by dropout captured the uncertainty inherent in the learned transition structure. While the model under deterministic conditions logically produced no variance over different entropy levels, the average variance across Monte Carlo dropout predictions increased with the entropy of the true transition distribution (see Figure \ref{fig:markov-chain-experiment}, bottom right). This demonstrates that Dropout Monte Carlo after \citep{gal2016dropout} can be used to quantify uncertainty in PBN networks.

\FloatBarrier

\section{Discussion}

\subsection{Implications for PC models}

Our results establish PBNs as a PC-compliant framework capable of learning brain aligned abstract representations that generalize to downstream tasks and have representational properties that conventional feedforward networks do not offer, including distinct brain-aligned top-down decoder representations, reproducing hallmarks of PC in biological brains \citep{schwiedrzik2017high}, and being capable of quantifying uncertainty. Together, these findings provide evidence that systems exclusively built around PC can support brain-aligned representations and computations extending to high levels of abstraction.

More broadly, PBNs provide a flexible unified framework in which PC models can be constructed, systematically constrained, and compared under controlled conditions. By training parameter-matched model variants that progressively impose different PC principles, the framework enables individual components of PC theory to be isolated and evaluated against one another. This provides neuroconnectionism with a framework for testing PC hypotheses at a level of representational complexity previously difficult to access.

\subsection{Prior subtraction and other operations}

Previous work has suggested that the explicit subtraction operation between bottom up information and top-down predictions assumed by classic PC may be fundamentally problematic for hierarchical representation learning \citep{rao2023predictive, lowet2024predictive}. Strict subtraction removes information that is explained away by top-down predictions at lower levels or earlier time points, implying that information about predictable stimuli should progressively disappear from deeper or later representations. This assumption is difficult to reconcile with evidence that highly predictable stimuli can nevertheless elicit robust, increasingly abstract representations throughout the ventral visual stream \citep{grill2014functional, hong2016explicit, kriegeskorte2008matching, isik2014dynamics}. Our model comparisons support this concern: PBN variants using subtraction generally performed worse than subtraction-free variants in all our comparisons. These results provide evidence that the subtraction operation of classic PC can impair the propagation of information required for hierarchical abstraction. These findings are consistent with proposals from PC literature that the interaction between top-down predictions and bottom-up signals may instead rely on mechanisms such as precision weighting, gating, or scaling, \citep{feldman2010attention, bastos2012canonical, gilbert2013top, de2011multiplicative, kanai2015cerebral}. The range of proposed mechanisms reflects an ongoing debate about how top-down predictions interact with bottom-up signals in cortical computation. Determining which mechanisms best capture these interactions is an important direction for future work. PBNs provide a flexible framework in which these alternatives can be easily implemented and compared under controlled and naturalistic conditions.

\subsection{Contrasting versus Predicting}

A specific feature of most PBN variants considered here is the use of contrastive predictive coding (CPC) \citep{oord2018representation}, which requires negative samples and thereby introduces dependencies between stimulus representations across time. This makes learning temporally non-local. \cite{oord2018representation} describe several variants of the CPC objective that contrast the true future sample with negative samples drawn from different temporal distances, thereby increasing or decreasing the temporal non-locality. One possible way to reconcile this dependency with biological constraints is to assume that temporally distant representations can be retained in memory and accessed locally during learning.

An alternative approach that omits temporal non-locality altogether is to learn representations using purely predictive or invariance-based objectives rather than contrasting against other samples, as used here in the PBN classic PC variant. Such objectives are theoretically capable of learning useful representations without explicit negative samples \citep{chen2021exploring, assran2025v} and may produce more brain-like representations and phenomenological correspondences \citep{gutlin2025predictive, gutlin2026predictive}. However, our current results do not support the scalability of this approach in the PBN framework. Methods such as SimSiam \citep{chen2021exploring}, BYOL \citep{grill2020bootstrap}, VICReg \citep{bardes2021vicreg}, or V-JEPA \citep{bardes2024v, assran2025v} provide possible approaches to solve this issue, but their suitability in hierarchical local learning contexts remains unresolved. Investigating PBN networks with different losses and comparing their performance to contrastive PC losses therefore represents a promising direction for future work.

\subsection{Insights for Deep Learning}

While the feasibility of local learning in DNNs has been investigated previously \citep{siddiqui2023blockwise}, we show that it extends to brain-like, rich hierarchical representations. This finding is striking because backpropagation provides the linearly optimal weight update in gradient descent, so any departure from backpropagation is sub-optimal in this sense. Optimization becomes harder in local learning where gradients do not flow across hierarchical levels, because there is no mechanism ensuring that weight updates at one hierarchical level will also benefit other levels. The fact that despite this additional constraint, locally restricted PC-inspired models are capable of learning rich representations, opens up new possibilities for investigating new types of learning systems. For example, local PC-based systems might have potential advantages, such as more resilience against catastrophic forgetting \citep{lee2022brain}, faster convergence (consistent with our observations during training), or potentially more biologically aligned learning trajectories, although this remains to be investigated systematically. This can be addressed in future work, facilitated by the framework introduced here.

\section{Conclusion}

In this work, we introduced Predictive Brain Networks (PBNs), a framework that combines the biological principles of predictive coding with the representational power of deep neural networks. We showed that PBNs learn rich, brain-aligned representations while capturing key signatures of predictive coding. Beyond providing a computational demonstration of the power of predictive coding, PBNs offer a flexible framework for systematically testing hypotheses about learning and information processing in the brain.

\section{Methods}

\subsection{Model architecture}
\label{sec:model-architecture}

The PBN framework is a hierarchical recurrent architecture composed of stacked autoencoders, with the overall architecture illustrated in Figure \ref{fig:general-overview}. Each hierarchy level is implemented as an independent autoencoder whose input is the compressed latent representation of the level below. Stacking these autoencoders forms a connected hierarchy of encoders and decoders that is trained as a single system, enabling progressively higher levels to learn increasingly abstract representations while retaining the local learning structure of each level. PBNs combines elements from autoencoders \citep{kingma2013auto, ronneberger2015u, sonderby2016ladder, chung2015recurrent}, contrastive learning \citep{oord2018representation, hinton2022forward, chen2020simple}, PC networks \citep{lotter2016deep, whittington2017approximation, millidge2024predictive}, teacher-student networks \citep{grill2020bootstrap, chen2021exploring}, localized learning \citep{siddiqui2023blockwise}, and self-supervised sequence learning models \citep{bardes2024v, assran2025v, tong2022videomae, zbontar2021barlow}. Each hierarchy level contains two skip connections. The posterior connection is a forward encoder-to-decoder connection analogous to the skip connection in a U-Net and provides the information required to form the posterior. The prior connection transmits the previous top-down decoder state to the encoder, where it is combined with the new bottom-up input to incorporate prior information. The prior connection makes the architecture inherently recurrent: the decoder state at each hierarchy level is retained and used as the prior for the subsequent encoder input. Because the decoder predicts the subsequent element in the sequence, this recurrence also makes PBNs inherently autoregressive. The basic recurrent state is necessarily tied to the decoder output. PBNs therefore optionally include a memory module in the posterior pathway to provide additional recurrent capacity and represent ongoing states that are not constrained to the form of the decoder output. The memory module is not required for the basic architecture. The architecture is modular, with individual components and operations replaceable according to the PC formulation being implemented. In this sense, the framework can represent a broad range of PC and Bayesian brain models while retaining the same hierarchical encoder-decoder structure.

\subsection{Loss}

Losses can be placed at any location in the network via loss layers, using the variables available at that location. In a PC setting, the most meaningful location is the prior fill-in layer, where the top-down decoder output from the previous time step can be compared with the bottom-up encoder input at the new time step. This comparison can be formulated either as an approximation objective (using MAE, MSE, or other invariance-based distance metrics) as in typical prediction-based PC approaches, or as a contrastive objective that promotes similarity between matching representations and separation between mismatched representations, as in CPC \citep{oord2018representation} or Forward-Forward \citep{hinton2022forward}. The posterior fill-in can likewise be used to define approximation or contrastive losses, although such losses are less directly tied to the PC formulation. Other network locations can provide meaningful sites for additional regularization; for example, losses applied at the end of the encoder can impose activity or covariance regularization to enforce PC-like sparsity assumptions or variational-autoencoder-like latent-prior assumptions. This flexibility allows the framework to implement a broad range of PC methods while retaining a common underlying architecture.

\subsection{Stop gradients}

The general understanding of the brain is not that processes approximating gradient descent and backpropagation \citep{rumelhart1986learning} are generally impossible, but rather that they become progressively less likely the longer the chain of nonlinear operations involved \citep{lillicrap2020backpropagation, lillicrap2019backpropagation, whittington2019theories, bartunov2018assessing}. We do not deny gradient descent or shallow backpropagation by default; rather, we aim to prevent biologically implausible gradient transmission through gigantic networks and the reliance on exactly calculated partial derivatives over thousands of nonlinear operations, which we consider beyond the capability of the brain. Accordingly, we use stop gradients to restrict gradient flow to single hierarchy levels. Stop gradients can be placed at any point in the network via stop-gradient layers, though the most meaningful variant is typically to place them between hierarchy levels. Different placements nonetheless allow for much more fine-grained control, for instance determining whether gradients are permitted to flow backwards through time within a single hierarchy level. Placement can likewise determine whether a hierarchy's encoder receives gradients from the immediately higher loss, that is, whether only the top-down prediction is tuned to approximate the bottom-up signal, or whether the bottom-up signal is tuned as well by the immediately adjacent top-down prediction; the latter works well for contrastive losses but can produce trivial solutions in pure invariance settings. Controlling the placement of stop gradients is thus one of the most important factors in setting the learning conditions of the model.

\subsection{Vision PBN implementation}

\begin{figure}[htbp]
    \centering
    \includegraphics[width=0.48\linewidth]{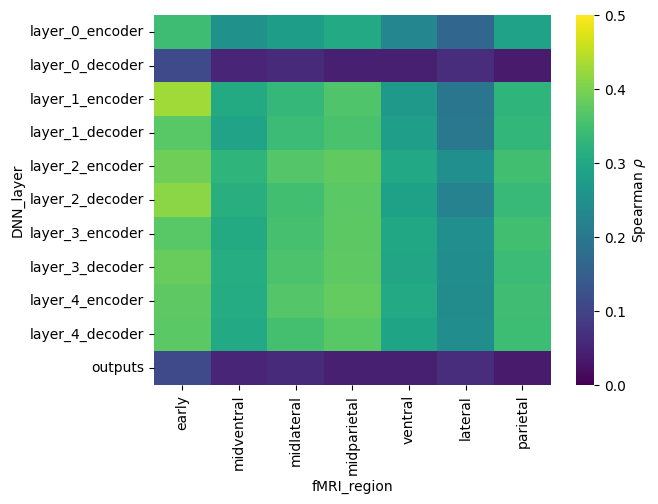}
    \hfill
    \includegraphics[width=0.48\linewidth]{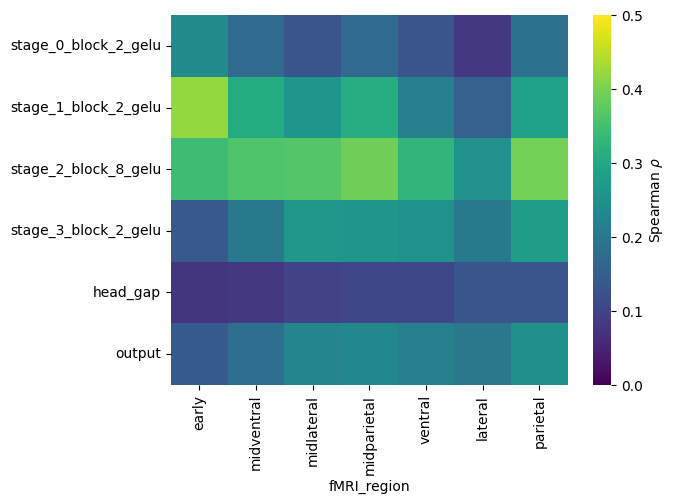}
    \vspace{0.5em}
    \includegraphics[width=0.48\linewidth]{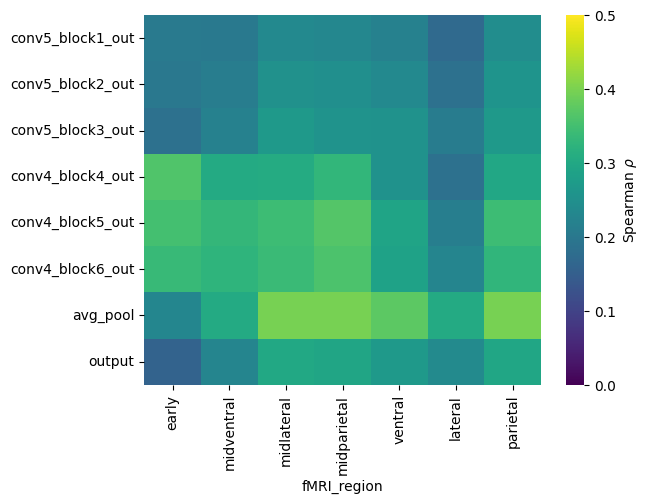}
    \hfill
    \includegraphics[width=0.48\linewidth]{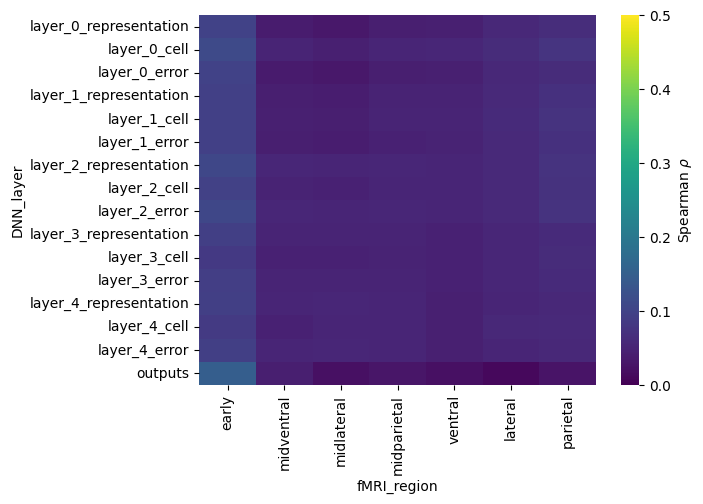}
    \caption{\textbf{Layerwise model-brain similarity.} Layerwise noise-corrected model-brain representational similarity, as a response to the Natural Scenes Dataset \citep{allen2022massive}, dissolved over model activations and brain regions. \textbf{Top left:} PBN local gradients, \textbf{Top right:} ConvNeXt, \textbf{Bottom left:} ResNet-50, \textbf{Bottom right:} PredNet.}
    \label{fig:layerwise_model_brain_similarity}
\end{figure}

For the implementation evaluated in this study, we implemented a PBN optimized for visual feature extraction from video time series. We adapted the PBN encoder to approximately match the stage-wise structure of ConvNeXt, using projection and downsampling layers to obtain corresponding spatial resolutions and channel dimensions and matching the number of residual convolutional blocks at each stage. The encoder comprises four convolutional stages with 96, 192, 384, and 768 channels, with progressively reduced spatial resolution; the first three stages each contained three residual convolutional blocks, while the fourth contained nine blocks arranged in three groups of three (evidence that learning occurred in all of those blocks for PBN local gradients is given in Figure \ref{fig:layerwise_model_brain_similarity}). The final convolutional stage is followed by global average pooling and two dense layers. Unlike the standard ConvNeXt implementation, which uses Layer Normalization, we used Batch Normalization, as this provided better performance with the CPC objective. To enable Monte Carlo dropout, we used standard dropout with a rate of 0.1 throughout the residual and dense layers rather than the ConvNeXt-specific branching/drop-path formulation. We designed the decoder to broadly mirror the encoder hierarchy, transforming the latent representation back to the input space using a minimal sequence of dense, upsampling, transposed-convolution, and skip-convolution layers. For all PBN models except the classic PC model, each hierarchy level incorporates a memory cell (Continuous Gated Cell) that projects the current input into a 32-dimensional memory state and uses a learned sigmoid gate to update this state from the previous memory and current input. The updated memory is projected back into the input space and added residually to the current activation.

\subsection{Model variants}
\label{sec:methods-model-variants}

\begin{figure}[htbp]
    \centering
    \includegraphics[width=0.32\linewidth]{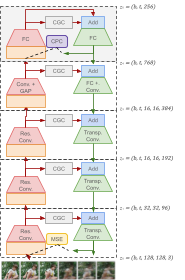}
    \hfill
    \includegraphics[width=0.32\linewidth]{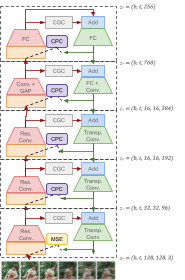}
    \hfill
    \includegraphics[width=0.32\linewidth]{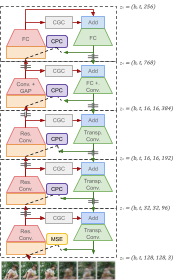}
    \vspace{0.5em}
    \includegraphics[width=0.32\linewidth]{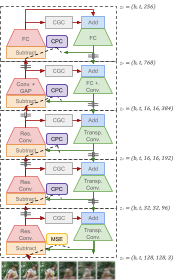}
    \hfill
    \includegraphics[width=0.32\linewidth]{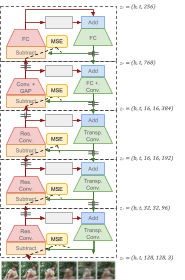}
    \hfill
    \makebox[0.32\linewidth]{}
    \caption{\textbf{Visualization of PBN model variants.} \textbf{Top left:} PBN backprop, \textbf{Top center:} PBN backprop + local loss, \textbf{Top right:} PBN local gradients, \textbf{Bottom left:} PBN local gradients + subtract, \textbf{Bottom right:} PBN classic PC. Legend: CGC: Continuous Gated Cell. Simple residual recurrent cell with a gated residual state. (\(\DashVDash\)): Stop gradients. At this location, gradients are prohibited from passing.}
    \label{fig:model_variant_architecture_visualization}
\end{figure}

With this general architecture, we built a range of models and varied aspects of the architecture, specifically the loss, and the gradient flow (see Figure \ref{fig:model_variant_architecture_visualization}). For the first four variants, we used the PC variant CPC (CPC, \citep{oord2018representation}), a contrastive objective that aligns with the principles of PC insofar as it can be framed as reducing the activational distance between the model's prediction and the true, matching future input, while maximizing the activational distance to unexpected, mismatching inputs. At the lowest level of the architecture, all implemented models use a simple mean squared error (MSE) loss for pixel-level reconstruction, as there is no need to compress information at the pixel level (CPC would still work at this level but yield qualitatively less interpretable reconstructions). This reconstruction objective enables us to visualize model predictions and estimate pixel-wise uncertainty using Monte Carlo dropout. Although pixel-perfect reconstruction is not the primary modeling objective and reconstructions are generally coarse, they provide an additional view of network behavior (see Figure \ref{fig:figure-a7-prediction-examples-for-the-pbn-local}). We further introduced low, naturalistically plausible augmentations to simulate natural noise or unsharpness (Gaussian smoothing), lighting changes (hue), and saccades (random crops, which we constrained to a lower and constant level to approximate saccadic movement \citep{wang2021use, thorat2025predicting}), omitting implausible augmentations like unconstrained random cropping with zoom or image flips. We found this set of light augmentations to be sufficiently naturalistic; However, while developing this study we found that similar effects can be achieved using mere masking (representing noisy retinal input).

Using these components, we defined six model variants (for a visual explanation, see Figure \ref{fig:model_variant_architecture_visualization}). First, \textbf{PBN untrained} is an untrained version of the same architecture as the trained PBN variants, providing an additional control that quantifies the variance accounted for by the architecture alone. Second, \textbf{PBN backprop} is a variant using a single CPC loss at the highest level together with the reconstruction loss, with gradients allowed to flow through the entire network. This model therefore behaves very similarly to a typical self-supervised model. Third, \textbf{PBN backprop + local loss} uses a CPC loss at each level of the hierarchy and thus formally complies with the definitions of PC above, but still allows gradients to flow through the entire network. Fourth, \textbf{PBN local gradients} uses a CPC loss at each level of the hierarchy and additionally blocks gradient flow between hierarchy levels, such that the gradients of each loss are restricted to a single level. Fifth, \textbf{PBN local gradients + subtract} uses a CPC loss at each level of the hierarchy, blocks gradient flow between hierarchy levels, and additionally introduces the classic PC operation of explicit subtraction between the prior and the incoming input, such that only the unexplained residual is propagated forward. This subtraction operation is not required to qualify as a PC model but constitutes a distinguishing feature of classic PC \citep{rao1999predictive}. Finally, \textbf{PBN classic PC} is a model variant designed to approximate classic PC as closely as possible: no contrastive loss, only a local MSE reconstruction loss at every level of the hierarchy; strictly local gradients, restricted within each hierarchy level; and no augmentation, as reference models trained with augmentation or masking did not outperform this unaugmented variant. Importantly, this variant also includes the subtraction operation characteristic of classic PC.

\subsection{Control models}

\subsubsection{ResNet-50 imagenet}

We used the standard supervised ImageNet-trained ResNet-50 model \citep{he2016deep} from keras.applications.ResNet50 \citep{chollet2015keras}, comprising 25,583,592 trainable parameters. ResNet-50 is widely used as a model in NeuroAI and neuroconnectionist research. For inference, each video frame was resized to the input dimensions required by the model (224 × 224 × 3) and scaled to the required input range (0–255). We extracted activations over time by presenting each video frame independently as an image and compiling the resulting activations across all network blocks and time steps.

\subsubsection{ConvNeXt-tiny imagenet}

ConvNeXt-tiny imagenet \citep{liu2022convnet} was chosen as a reference model for producing highest brain alignment in the Brain-score vision/neural benchmark \citep{schrimpf2018brain} at the time of the conception of this project (retrieved 2025-01-15) and for being easily accessible via keras.applications.ConvNeXtTiny \citep{chollet2015keras}, comprising 28,589,128 trainable parameters. For inference, each video frame was resized to the input dimensions required by the model (224 × 224 × 3) and scaled to the required input range (0–255). We extracted activations over time by presenting each video frame independently as an image and compiling the resulting activations across all network blocks and time steps.

\subsubsection{PredNet}

We used the original PredNet implementation from \citep{lotter2016deep}, written in Keras, adapting it to our Keras version while keeping the operations equivalent. We used a similar filter ratio to that of the original paper, with the latent-state filter count scaling by a factor of two across layers (32, 64, 128, 256, and 512), while adapting the number of layers such that the PredNet model contains a similar number of parameters to the PBN models (33,726,563 versus 34,610,952). This resulted in a five-layer PredNet, matching the number of hierarchy layers in the PBN models. PredNet was trained on the same dataset as the PBN models. We used the same normalization as in the original paper, scaling inputs to [0, 1], and used an input size of 128 × 128 × 3, which is the closest match to the 128 × 160 × 3 input used in the original paper. The original paper primarily reports two loss settings: \(L_{\mathrm{0}}\), in which only the lowest-layer loss is weighted (weight 1) and all other layer losses are set to zero, yielding better reconstruction but worse classification performance, and \(L_{\mathrm{all}}\), in which the lowest-layer loss is weighted by 1 and all other layer losses by 0.1, yielding slightly worse reconstruction but better classification performance. We used the \(L_{\mathrm{all}}\) setting because we are interested in the abstract representations formed by the model.

\subsection{Training}
\label{sec:methods-training}

\begin{figure}[htbp]
    \centering
    \includegraphics[width=0.32\linewidth]{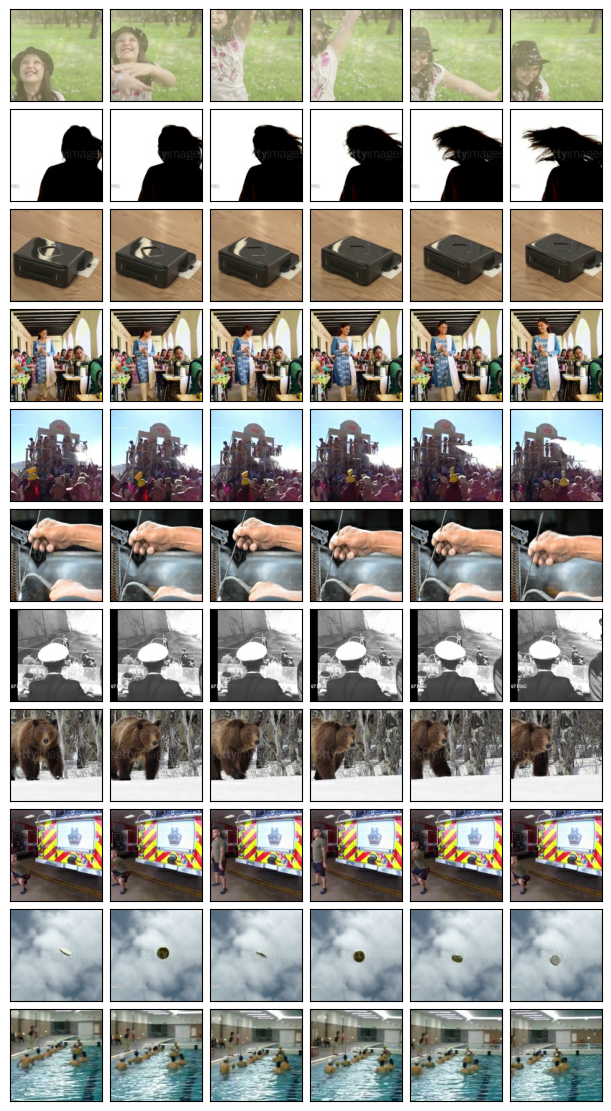}
    \hfill
    \includegraphics[width=0.32\linewidth]{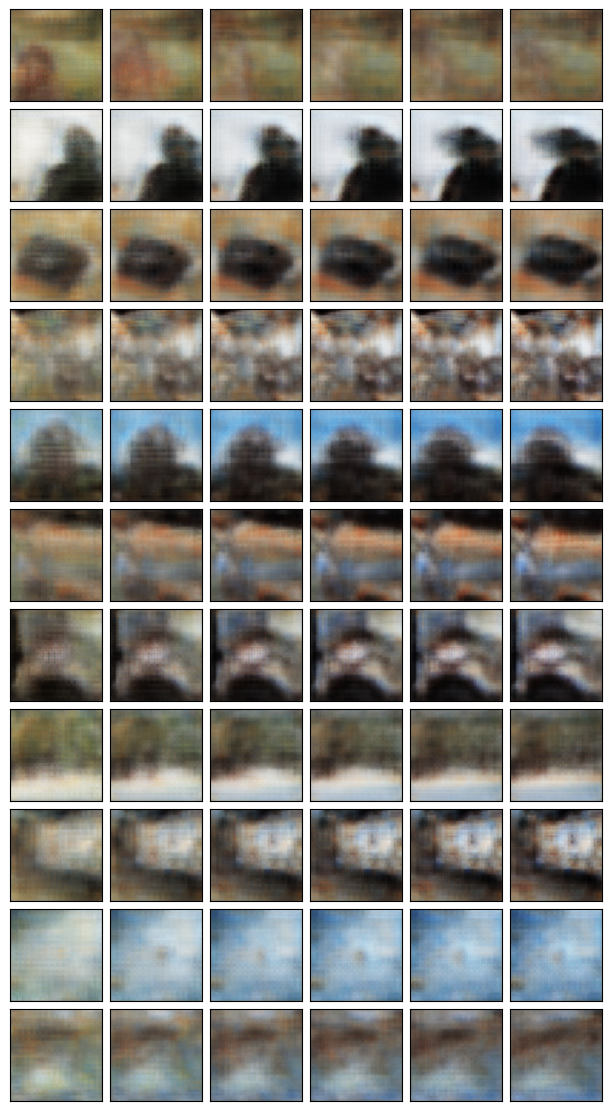}
    \hfill
    \includegraphics[width=0.32\linewidth]{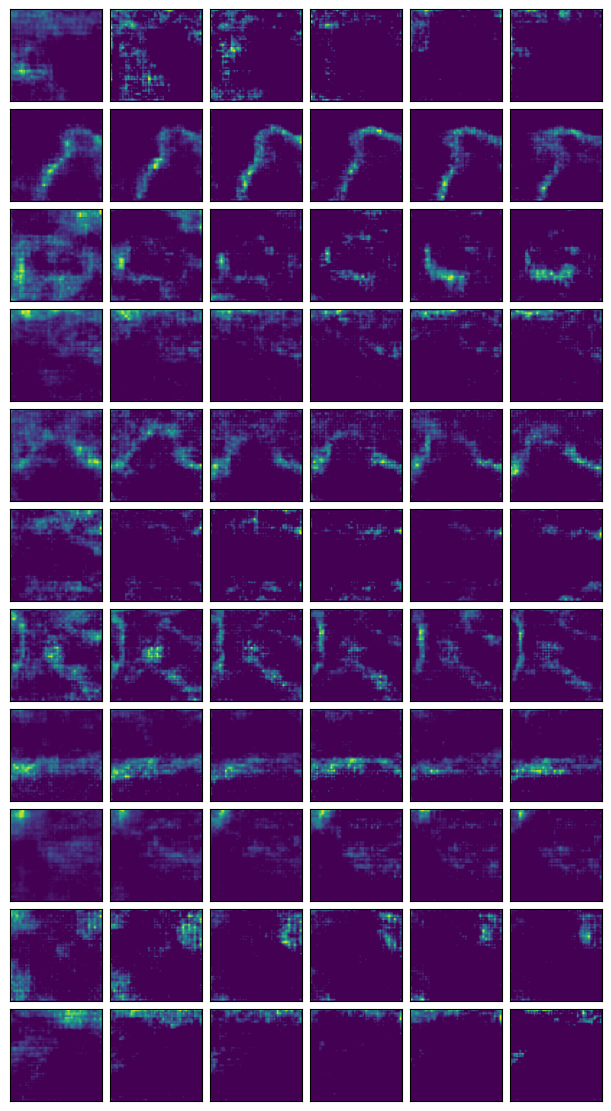}
    \caption{\textbf{Prediction examples.} Prediction examples for the PBN local gradients model variant. \textbf{Left:} Example input series from the test dataset. \textbf{Center:} Corresponding model predictions from the lowest-level decoder layer. \textbf{Right:} Model-inferred spatial uncertainty estimated using Monte Carlo dropout. Reconstructions are generally coarse, as pixel-perfect reconstruction is not the primary modeling objective (see \hyperref[sec:methods-model-variants]{Methods}).}
    \label{fig:figure-a7-prediction-examples-for-the-pbn-local}
\end{figure}

We trained all relevant models (PredNet and all PBN models except for the untrained variant) on short naturalistic video clips from \citep{monfort2019moments, gifford2024algonauts, dave2020tao}. We used the test split of the MIT dataset as the held-out test set. Videos ranged from 1 s to over 2 h in duration, with most videos being approximately 3 s long; videos longer than 3 s were segmented into clips of at most 3 s. Although the models can process sequences of arbitrary temporal length, training was performed on sequences of 6 frames sampled at 3.2 Hz. Models were implemented using Keras 3 \citep{chollet2015keras} (making all models work with Pytorch, TensorFlow, or JAX backends) and trained on a TensorFlow backend \citep{abadi2015tensorflow}. Training used stochastic gradient descent (SGD) with momentum and cosine learning-rate decay, with an initial learning rate of 0.075 decayed to zero over approximately 2.6 million training steps, momentum of 0.9, weight decay of 1e-6, and global gradient clipping at a norm of 1. During training, the training data were shuffled after each batch. Models were trained using an early-stopping/model-selection procedure, retaining the checkpoint with the lowest validation loss. Training continued for at least 350 epochs and was terminated when validation performance failed to improve for 50 epochs; all trained models converged before reaching the minimum training duration. Post-training test examples, including inputs, model predictions, and uncertainty estimates, for the PBN local-gradients variant, are shown in Figure \ref{fig:figure-a7-prediction-examples-for-the-pbn-local}.

\subsection{Representational similarity analysis}

We evaluated whether PBN models match commonly used models in terms of brain similarity using representational similarity analysis (RSA; \citep{kriegeskorte2008representational}). For this, we used stimulus images and average fMRI responses from the Natural Scenes Dataset (NSD; \citep{allen2022massive}), restricted to the 515 samples shared across all subjects. Images were reshaped to (128, 128, 3) and repeated across three time steps as a still-image video to allow the recurrent models a warmup period. Human RDMs were constructed by taking the average fMRI response to each image, filtered by region (early, midventral, midlateral, midparietal, ventral, lateral, parietal; following a region definition from \citep{thorat2025predicting, allen2022massive}), and calculating the pairwise correlational distance (1 - rho; scipy.spatial.distance.pdist) between images, yielding a set of RDMs for each region and subject. Model RDMs were extracted by feeding the same images into each model and extracting activations after each processing stage (each hierarchy level for PBN, each block and average pooling stage for ResNet-50 and ConvNeXt, and each layer for PredNet), yielding a set of RDMs for each time step and layer of each model. We normalized each correlation by the estimated lower-bound RSA noise ceiling following \citep{nili2014toolbox}, calculated for each region as the mean leave-one-subject-out correlation with the average RDM of the remaining subjects. To ensure that these results did not simply reflect selection of maximally favorable time steps or layers, we used a reweighted RSA approach. For each subject and region, we fitted each layer and time step of each model on external, non-515 NSD images, drawing 32 samples and fitting the model to predict the brain region over 200 separate draws, following the approach outlined in \citep{storrs2021diverse}. The resulting betas for each subject, region, layer, and time step were then applied to the held-out set of 515 test images to generate a reweighted RDM for each model, which was Spearman correlated to each region of each subject. Pairwise comparisons between all model combinations (reported in Figure \ref{fig:rsa_reweighted_pairwise}) were performed using Wilcoxon paired-samples tests (\citep{wilcoxon1945individual}) with false discovery rate correction following the Benjamini-Hochberg procedure (\citep{benjamini1995controlling}).

\begin{figure}[htbp]
    \centering
    \includegraphics[width=\linewidth]{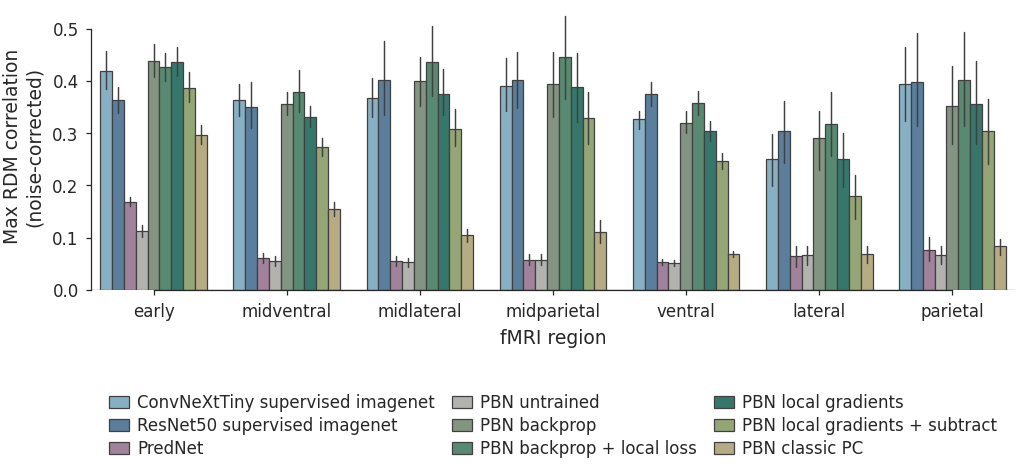}
    \vspace{0.5em}
    \includegraphics[width=\linewidth]{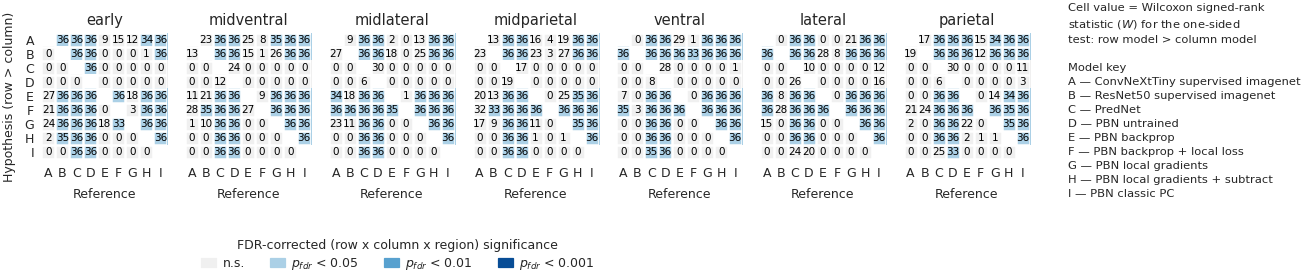}
    \caption{\textbf{Non-reweighted representational similarity analysis (RSA).} \textbf{Top:} Noise-ceiling-corrected correlations between each model’s maximum layer-wise correlation and responses in different brain regions to natural-scene images. \textbf{Bottom:} Pairwise post-hoc comparisons between models for non-reweighted RSA with cross-validation. Models are labelled as follows: A, ConvNeXt-Tiny (supervised ImageNet); B, ResNet-50 (supervised ImageNet); C, PredNet; D, PBN untrained; E, PBN backprop; F, PBN backprop + local loss; G, PBN local gradients; H, PBN local gradients + subtract; and I, PBN classic PC. Each cell shows a one-sided Wilcoxon test of whether the row model outperforms the column model. Numbers indicate the Wilcoxon test statistic, W, and background colours indicate significance levels after false discovery rate (FDR) correction across models, columns, and brain regions.}
    \label{fig:rsa_maximum}
\end{figure}

To confirm our results, we performed an alternative approach, where we simply calculated the maximum Spearman correlation across layers and time steps, yielding one maximum model-region correlation per subject. This additional analysis yielded qualitatively similar results. Its subject-average Spearman correlation and spread, together with pairwise comparisons between model variants, are shown in Figure \ref{fig:rsa_maximum}.

\subsection{Unique decoder variance}

To determine whether the decoder explains variance in human brain responses that is not already explained by the encoder, we quantified the variance uniquely attributable to decoder representations. For each PBN model, subject, and visual region, we extracted all encoder and decoder activations for the NSD \citep{allen2022massive} 515 shared images as described above and fitted linear regression models using all layers and time steps. We first fitted a model containing the encoder and decoder representations jointly and then a nested model containing encoder representations alone. The unique variance explained by the decoder was defined as the incremental coefficient of determination, \(R^2(\text{brain} \mid \text{decoder}) = R^2(\text{brain} \mid \text{all encoder layers} \cup \text{all decoder layers}) - R^2(\text{brain} \mid \text{all encoder layers})\). We corrected the resulting values for the natural noise ceiling by dividing the explained variance by the squared noise-ceiling correlation mentioned above \citep{nili2014toolbox}. We then compared models with one another using post-hoc pairwise tests, performed separately for each visual region. For every ordered pair of models, we tested whether the row model explained more unique decoder variance than the column model, using one-sided paired-samples Wilcoxon signed-rank tests \citep{wilcoxon1945individual}, paired by subject. This yielded a full matrix of directional comparisons per region, in which the direction of an effect is given by the position of the cell (row > column versus column > row). Significance levels were corrected across all regions and ordered model pairs using the Benjamini-Hochberg \citep{benjamini1995controlling} false discovery rate procedure ( \((\alpha = 0.05)\) ).

\subsection{Latent representations readout performance}
\label{sec:methods-readout-performance}

To evaluate general readout performance, we used two action-recognition video datasets: the test partition of Moments in Time \citep{monfort2019moments} and the UCF-101  \citep{soomro2012ucf101} benchmark, which is commonly used but easier to solve. Videos were converted into fixed-length clips matching the training clips in frame size, length, and frame rate. For each dataset, we extracted latent representations from each evaluated model using the deepest encoder state available. We evaluated the extracted representations using two lightweight downstream readouts, a 1-nearest-neighbor (kNN-1) classifier and a logistic-regression classifier, each using 5-fold cross-validation. For each dataset and model, we report the mean and standard deviation across cross-validation splits.

\subsection{Replicating neural effects of prediction}
\label{sec:methods-replication}

To test whether PBNs can reproduce neuroscientific findings that are difficult to model with established deterministic DNNs, we evaluated whether it reproduces the effects reported in \citep{schwiedrzik2017high}, in which lower levels of macaque visual cortex inherit prediction signals from higher-level, identity-selective processing stages. We closely followed their violation-of-expectation paradigm, in which macaques learn to predict a specific face image following another face image, establishing expectations over both identity and viewpoint. We used the face stimuli from \citep{thomaz2010new}, which provide the same relevant stimulus categories as \citep{schwiedrzik2017high} while offering a larger, higher-quality, and more systematically organized stimulus set. Predictor and successor identities were paired one-to-one across two non-overlapping groups of face identities, with face rotations of -90°, 0°, and 90°. Viewpoint transitions were balanced across identity pairs to prevent systematic viewpoint biases. The resulting predictor-successor pairs were used to train the network for 100 epochs.

After training, we evaluated four variants: training, view violation, identity violation, and combined view + identity violation. As in the original paradigm, the successor image was physically identical across all variants, with only the predictor image manipulated, ensuring that differences in the response to the subsequent stimulus could not be attributed to differences in its low-level image statistics. We passed all variants through the fine-tuned PBN and extracted latent encoder and decoder states from the first two hierarchy levels, theoretically best corresponding to the lowest level used in the original study. Prediction error was quantified as the difference between the encoder latent and the corresponding decoder prediction. For each condition, prediction error was expressed as the percentage change relative to the trained-condition baseline, \(\frac{\lvert PE_{\mathrm{condition}} - PE_{\mathrm{trained}}\rvert}{\lvert PE_{\mathrm{trained}}\rvert} \times 100\). We evaluated the first three time steps after the stimulus switch, defining the first as the early window and the second and third as the late window. To test whether identity violations produced a stronger temporal increase in prediction-error change than viewpoint violations, we calculated the early-to-late change for each condition and compared these changes using one-sided paired-samples t-tests \citep{student1908probable}.

\subsection{Next frame prediction}

\begin{figure}[htbp]
  \centering
  \includegraphics[width=0.5\linewidth]{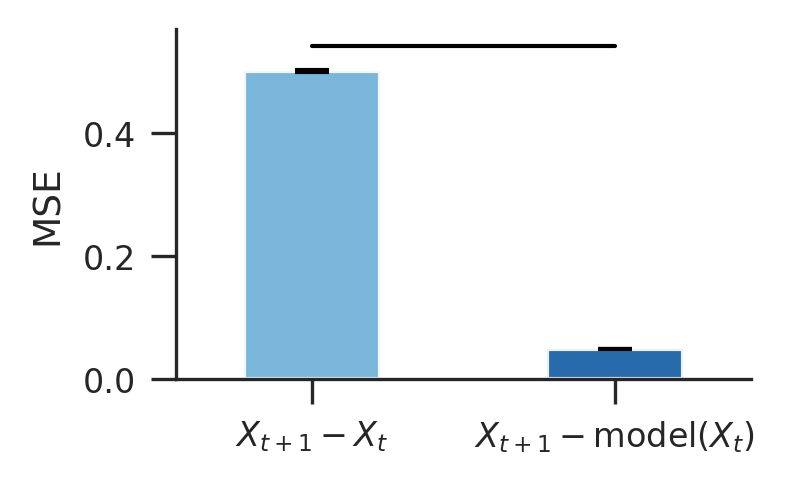}
  \caption{\textbf{Future prediction on the Markov dataset.} To test whether PBN predicts stimulus-sequence dynamics rather than merely extracting abstract features from individual inputs, we evaluated next-state predictions on the test set against a persistence baseline that assumes the current state remains unchanged, \(X_{t+1} = X_t\). The mean squared error (MSE) between the ground-truth future state and the model prediction, \(\mathrm{MSE}(X_{t+1}, \mathrm{model}(X_t))\), was significantly lower than that between the ground-truth future and current states, \(\mathrm{MSE}(X_{t+1}, X_t)\) (Wilcoxon \(W = 986037888,\ p < 10^{-37}\) ). These results indicate that PBN predicts future states more accurately than simply reproducing the current state.}
  \label{fig:figure-a5-future-prediction-on-the-markov-datase}
\end{figure}

We also used the Markov dataset and fine-tuned model to test whether the model was capable of predicting future states rather than merely encoding the current state (reported in Figure \ref{fig:figure-a5-future-prediction-on-the-markov-datase}). Specifically, we tested whether the fine-tuned model output was closer to the future state than the current state itself, i.e., whether \( \mathrm{MSE}(X_{t+1}, \mathrm{model}(X_t)) < \mathrm{MSE}(X_{t+1}, X_t) \). This was evaluated using a Wilcoxon signed-rank test.

\subsection{Uncertainty quantification}
\label{sec:methods-uncertainty-quantification}

We used a Markov chain experiment to investigate whether the model could 1. approximate the transition uncertainty of a known Markov process and 2. reflect differences in Markov-state uncertainty in the variance of its predictions under Monte Carlo dropout. We constructed a visual Markov chain in which each input frame was a black image containing one of four possible white-corner states, corresponding to four discrete Markov states: top left (A), top right (B), bottom left (C), and bottom right (D) (see Figure \ref{fig:markov-chain-experiment}, top). The state in the current frame determined the probability distribution over states in the subsequent frame according to a predefined 4x4 Markov transition matrix, which served as the ground-truth transition model (see Figure \ref{fig:markov-chain-experiment}, bottom left). Overall, 1152 training and 1152 validation transitions were generated from this process, in which the temporal structure of the visual input was fully determined by the predefined transition matrix. We fine-tuned the PBN local gradients model on these sequences for 100 epochs.

To quantify convergence toward the true Markov dynamics, we computed the mean squared error (MSE) between the model-estimated and ground-truth transition matrices at each epoch and tested whether the final model-inferred matrix had lower error than the initial matrix using a paired Wilcoxon signed-rank test. To determine whether predictive variance reflected the uncertainty inherent in the Markov process, we quantified the uncertainty of each true transition distribution using Shannon entropy \citep{shannon1948mathematical}, calculated from each row of the ground-truth transition matrix. We then compared the entropy of each state with the variance of the model-inferred predictions with and without Dropout Monte Carlo. We fit an ordinary least-squares linear regressions between transition entropy and aggregated predictive variance separately for the dropout and no-dropout conditions to show that the Dropout Monte Carlo condition correlated strongly with the true transition entropy, while the non-dropout condition did not.

\section{Data availability statement}

The source code and pretrained models used in this study will be made publicly available upon publication.



\section{Acknowledgments}

We thank Johannes Singer (Freie Universität Berlin), Alexander Lenders (Max Planck School of Cognition, Leipzig), Stefan Balle (Freie Universität Berlin), and Radoslav Cichy (Freie Universität Berlin) and his lab for discussion, comments, and valuable feedback on the project. We thank Mahdi Enan (Universiteit Maastricht), Denise Kittelmann (Universiteit Maastricht) for fruitful discussion and comments in the conceptualization phase. This work was performed using the Curta HPC cluster at Freie Universität Berlin \citep{bennett2020curta}.

\section{Statement of Authorship and Verification}

The authors affirm that all scientific claims, design, data analysis, and raw text originate from the authors. Advanced automated textual processing software (specifically Anthropic Claude, OpenAI ChatGPT) was utilized exclusively for light code and light text editing. The entire manuscript and code base was reviewed by the authors, who maintain sole authorship and responsibility for the final text.

\bibliography{bibliography}

\section{Supplementary Material}

\end{document}